# Boarding House Renting Price Prediction Using Deep Neural Network Regression on Mobile Apps

**Malik Abdul Aziz, Fahmi Nurrahim, Prastyo Eko Susanto, Yurio Windiatmoko**

malik.aziz@students.uii.ac.id, fahmi.nurrahim@students.uii.ac.id, prastyo.susanto@students.uii.ac.id
yurio.windiatmoko@students.uii.ac.id

Unofficial Star, Yogyakarta, Indonesia

**Abstract**. Boarding house is the most important requirement, especially for college students who live far away from the city, place of his origin or house. However, the problem we see now is the uneven distribution of study places in Indonesia which 75% of the best top educational institutions come from the island of Java. So, students who are looking for boarding houses rent requires more effort in comparing the various aspects desired. They need to survey one by one to the boarding house they want, even though they can survey online, it still requires more effort to pay attention to the desired facilities one by one. Therefore, we then created an Mobile Application that can predict prices based on student needs by comparing several variables, namely city, area, type of boarding house, and facilities. So, students can easily estimate the ideal price. The results of this study prove that we have succeeded in predicting prices for boarding houses rent well based on the variables we have determined, and modeling that variables using Deep Neural Network Regression.

## 1. Introduction

A Boarding house or 'kos-kosan' in Indonesian is a room that can be lived in for a certain period of time and requires payment. Boarding house is the most important requirement, especially for college students who live far away from the city, place of his origin or house. However, the problem we see now is the uneven distribution of study places in Indonesia, which are concentrated in several points such as Java and certain points with varying quality of campuses. As stated by uniRank, in 2020 around 75% of the best top educational institutions come from the island of Java. Seeing the uneven distribution of the best highest education institutions, prospective students who want to study in the best places need to be far from their homes, to enable them study at their dream campuses, and of course most of them need boarding or temporary housing rent to live while they are migrants. Those who become migrants are faced with the problem of ignorance about the price of finding a boarding house rent considering the rent price changes every year [1].

In addition, looking for boarding houses rent requires more effort in comparing the various aspects desired. They need to survey one by one to the boarding house they want, even though they can survey online, it still requires more effort to pay attention to the desired facilities one by one. Besides having to compare boarding house one by one with another boarding house, they also need to put more effort for finding boarding houses in the case of online boarding surveys, considering the application platform for finding boarding houses is too much (such as: mamikos, kosKost, infokost, kost hunt, and so on) even in the same application it will be difficult to compare them because you need to pay attention and remember one by one the fees with different facilities and different prices. So the decision to choose a boarding house can be quite confusing, because people will be careful in determining their budget [2].

After being explained above, about the problems of finding a boarding house, we made a tool in the form of a mobile application, where the finders can compare all the available aspects of the boarding house and get the best price recommendation according to the desired aspect, so you can be sure with certainty, actually the boarding house renting price, what kind of the ideal suits by them.

## 2. Literature Review

There are several factors that influence the price of a boarding house rent. In a study by [3] they divided these factors into three groups, namely physical condition, concept, and location. Physical conditions are equipment that is owned by a house and can be observed by the human senses, including size, availability of a kitchen, garage, and age of the house [4]. Also according to [5] which concludes that there are several factors that affect students preferences, choose a boarding house. These factors can be summarized into six main factors. Therefore, these factors are the main variables : price, location, facility, security, environment, neighborhood, and building condition. (So that we try to simplify all of these factors into just city, area, type of kost, and facilities as our independent variables to price as dependent variables.)

Research conducted by [2], namely comparing several regression techniques such as multiple linear regression, ridge regression, LASS regression, elastic net regression, boosting regression, and gradient boosting regression in case studies of house price predictions. Where the results of the study were the gradient boosting algorithm had the highest accuracy score compared to other algorithms.

[6] have implemented a mobile application that can provide users with a novel way to view future housing price forecasts in London which considering the performance of the application, they separate the application to the client and server. Their research actually wasn't boarding house rent prediction, but it has the same contexts as regression mobile application which we develop in this research.

After describing some of these studies, the mobile applications for boarding rental price predictions have not been found yet. So that, we combine some of these research descriptions into a mobile application development research for boarding house rent price predictions by simplifying the feature into city, area, type of kost, and facilities as our independent variables compare with price as dependent variables. We also make the application not separated into client and server side, but embedding the model into mobile apps so that implementation can be done offline.

## 3. Data and Methods

In the implementation method to build a machine learning-based Getkos application as a cost predictor, the data must first be collected as a training data capital to build a cost price predictor model. The whole processes in this research and development are following the diagram below.

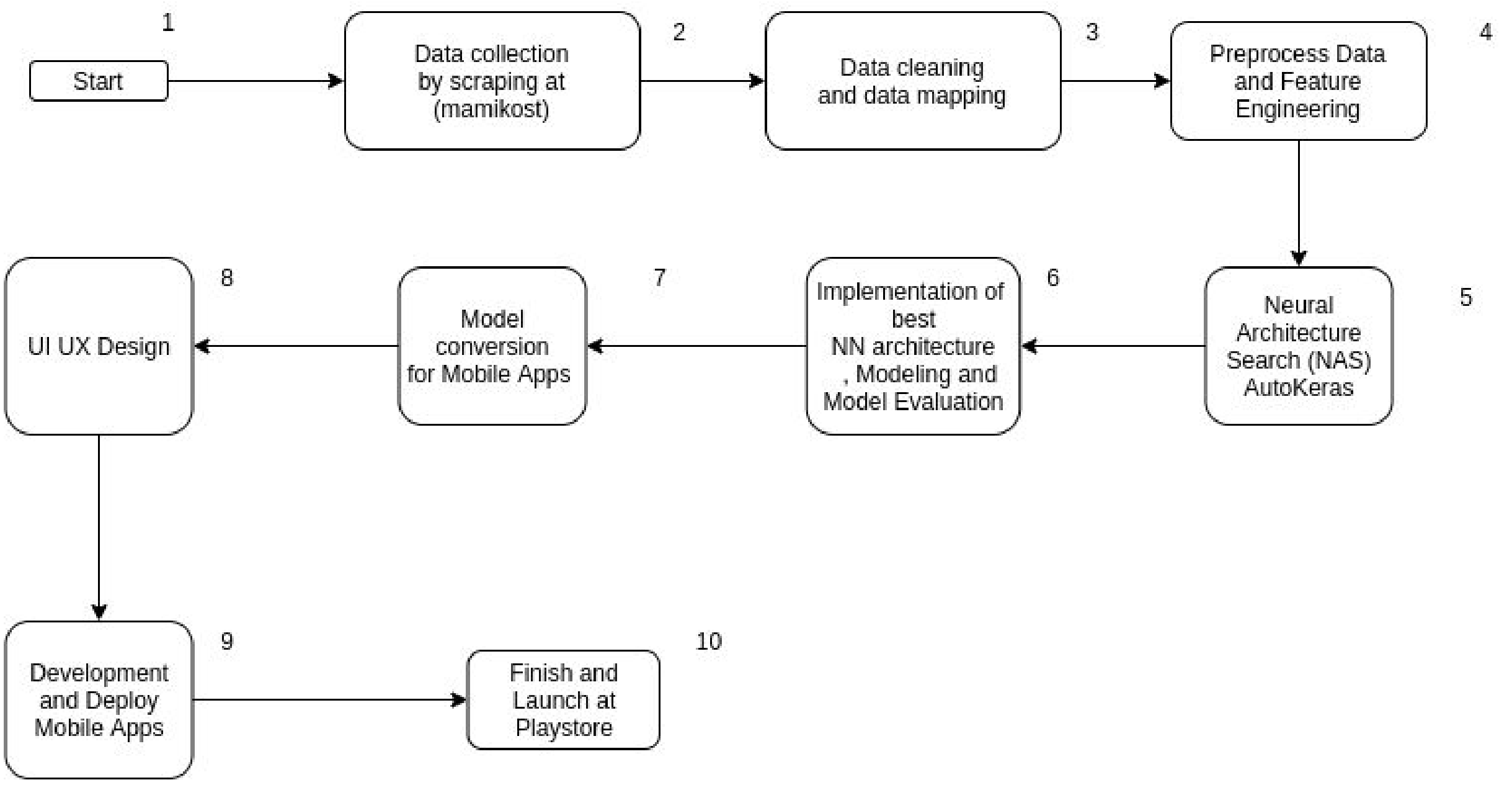

Figure 1. Workflow Diagram Processes of Research

## 3.1 Scraping Data

Data from various boarding rent prices are obtained by scraping the mamikos.com web. The data taken include the type of boarding house (male, female, and mixed), name of boarding house, location and area of boarding house, facilities, and price (per month). We use web scrapers tools to get the data. The web scraper tools are *Data miner pro* and *Web Scraper*.

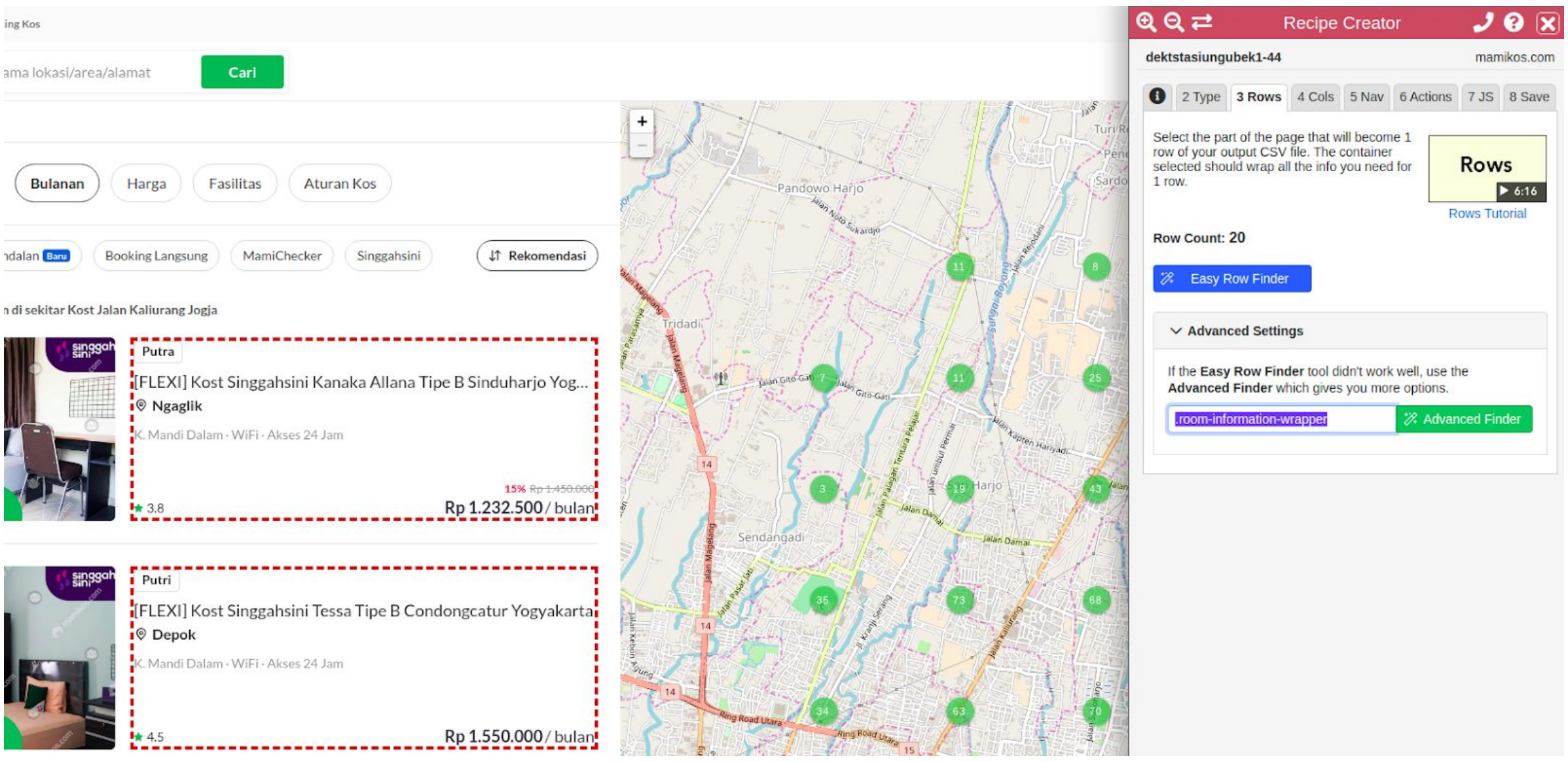

Figure 2. Data Miner Pro Scrapping Process

## 3.2 Descriptive Analytics

Several variables we used as a basis in providing for boarding house renting price predictions. These variables are city, area, boarding house type, and facilities. It is also called as independent variables in regression tasks and obviously rent price as the dependent variable.

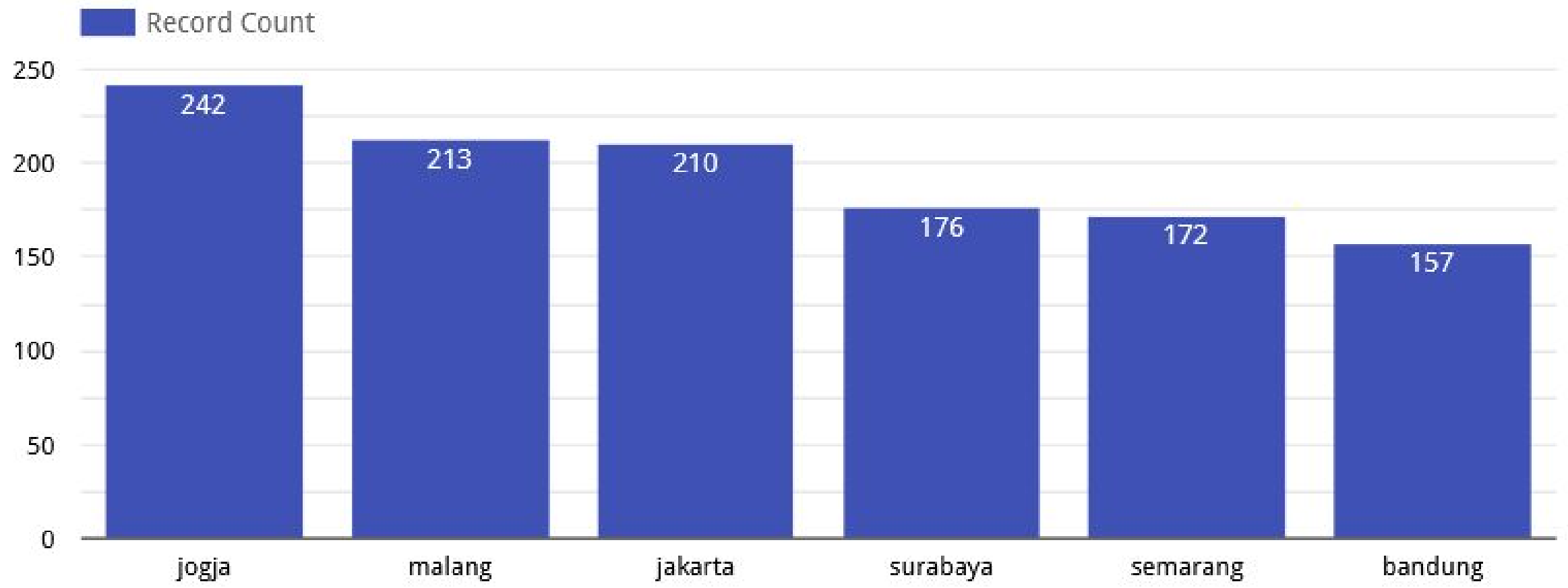

Figure 3. Distribution of City

Figure 3 shows the cities that we determined to get boarding rent data, namely Jogja, Malang, Jakarta, Surabaya, Semarang, and Bandung. The total count of data is 1,205 boarding houses.

| | kota | area |
|---|---|---|
| 1. | jogja | 23 |
| 2. | malang | 7 |
| 3. | jakarta | 24 |
| 4. | surabaya | 23 |
| 5. | semarang | 15 |
| 6. | bandung | 29 |

Figure 4. area count on each city

| | type_kos | Record Count ▾ |
|---|---|---|
| 1. | Campur | 511 |
| 2. | Putri | 441 |
| 3. | Putra | 214 |

Figure 5. type of boarding count

The area has a total count of 121 areas which are divided into 6 cities that have been mentioned above, which can be seen in figure 4. Then for the type of boarding house it is divided into 5 types, namely mixed, male, and female which can be seen in the distribution of the numbers in the figure 4.

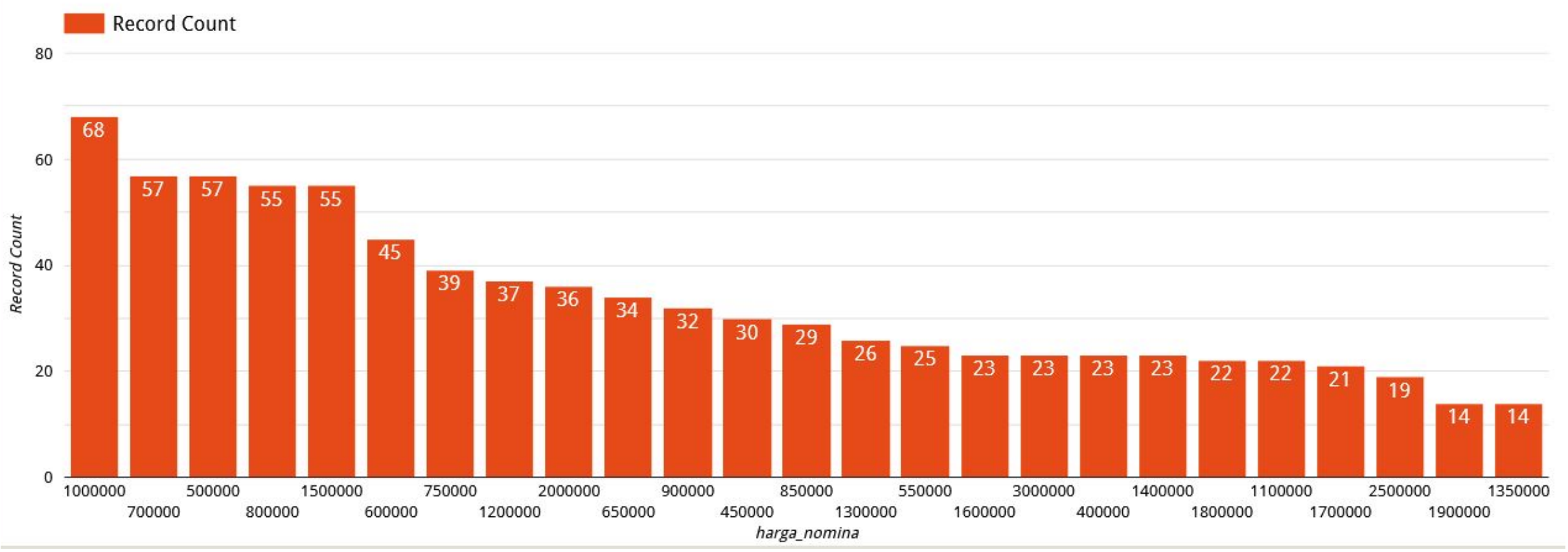

Figure 6. Record count of ranking price

In figure 6 is the order of top 25 from largest amount of price data, which costs Rp. 1.000.000 / month is the top quantity distribution with a total of 68 boarding houses.

## 3.3 Cleansing and Preprocessing Data

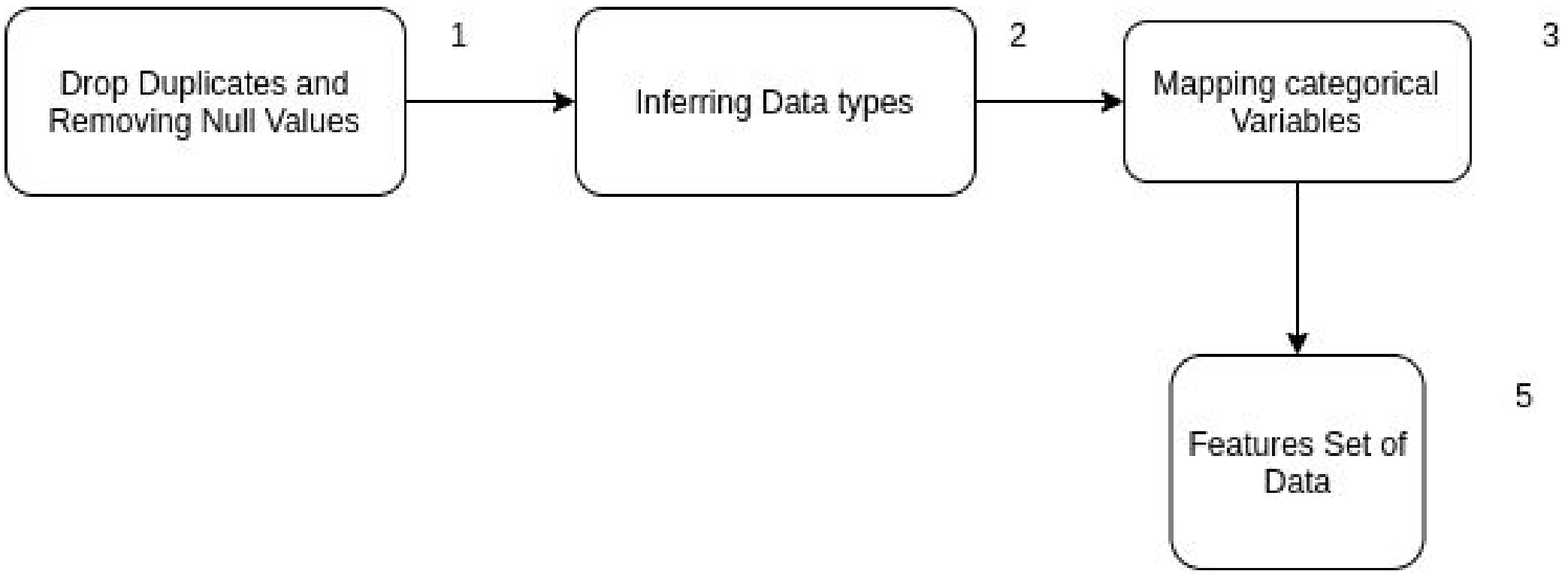

Figure 7. Data Cleansing and Preprocess Flow

Figure above are the flow of our data cleansing and preprocess, first after we already got all the data from web scraping, we need to drop duplicates some of the data, it because sometimes when we scrap from an area of a city it will be overlapped to other area within the same city. Then we remove null values when a categorical variable from the data is null.

Tables 1. Raw Data After Cleansing

| no | kost_name | kota | type_kos | area | facility_score | harga_nomina |
|---|---|---|---|---|---|---|
| **0** | Kost Pondok Nirwana Merr Tipe C Surabaya | surabaya | putri | rungkut | 6 | 1600000 |
| **1** | Kost Pondok Nirwana Merr Tipe B Surabaya | surabaya | putri | rungkut | 4 | 1500000 |
| **2** | Kost Pondok Nirwana Merr Tipe A Surabaya | surabaya | putri | rungkut | 4 | 1400000 |
| **...** | ... | ... | ... | ... | ... | ... |
| **1204** | Kost Cozy Putra Lowokwaru Malang | malang | putra | lowokwaru | 6 | 1400000 |

Finally after data has already been cleaned then we mapping or encode all the categorical variables into integer categorical value.

Tables 2. Data Features Ready for Train and Modeling

| no | kota | type_kos | area | facility_score | harga_nomina |
|---|---|---|---|---|---|
| **0** | surabaya | putri | rungkut | 6 | 1600000 |
| **1** | surabaya | putri | rungkut | 4 | 1500000 |
| **...** | ... | ... | ... | ... | ... |
| **1203** | malang | putri | lowokwaru | 4 | 550000 |
| **1204** | malang | putra | lowokwaru | 6 | 1400000 |

### 3.4 Neural Architecture Search with Autokeras

Neural architecture search (NAS) has been proposed to automatically adjust our deep neural networks. AutoKeras uses network morphology to maintain the function of the neural network while changing the structure of the neural network. Through more effective training during the search process, NAS can be helpful. This novel framework enables Bayesian optimization to guide the network morphology search for effective neural architecture [7]. The framework developed a neural network kernel and tree-like acquisition function optimization algorithm to effectively explore the search space.

````
```python
search =
StructuredDataRegressor(loss='mean_absolute_error',directory='models_sa
ved_data/')

# perform the search
search.fit(x=X_train, y=y_train, verbose=0)
```
````

Figure 8. Performing NAS with Autokeras in Regression Task

### 3.5 Modeling and Evaluation

After all the Neural architecture search (NAS) process finished in AutoKeras and the parameter has been tuned, we get our best architecture like this table below.

Tables 3. Best NN Regressor Architecture from NAS

| Layer (type) | Output Shape | Param |
|---|---|---|
| input_1 (InputLayer) | [(None, 4)] | 0 |
| multi_category_encoding | (None, 4) | 0 |
| normalization | (None, 4) | 9 |
| dense (Dense) | (None, 256) | 1280 |
| re_lu (ReLU) | (None, 256) | 0 |
| dense_1 (Dense) | (None, 512) | 131584 |
| re_lu_1 (ReLU) | (None, 512) | 0 |
| dense_2 (Dense) | (None, 128) | 65664 |
| re_lu_2 (ReLU) | (None, 128) | 0 |
| regression_head_1 (Dense) | (None, 1) | 129 |
| **Total params** | **198,666** | |
| **Trainable params** | **198,657** | |
| **Non-trainable params** | **9** | |

````
```python
# evaluate the model
mae, _ = search.evaluate(X_test, y_test, verbose=0)
print('MAE: %.3f' % mae)

>> MAE: 429086.375
```
````

Figure 9. MAE Evaluation from the Best Model

### 3.7 Konversi Model Tf-lite

The best model results that we obtained are then converted into tflite format to simplify the process of embedding the model into a mobile application for offline implementation.

````
```python
converter =
tf.lite.TFLiteConverter.from_saved_model('getkos_model')
tflite_model = converter.convert()

with open('model_getkos_regression.tflite', 'wb') as f:
````
````

````
    f.write(tflite_model)
```
````

Figure 10. Model Conversion to tflite

### 3.8 User Interface and User Experience Application

Below is the display design of the GetKos application as shown in Figure 8.

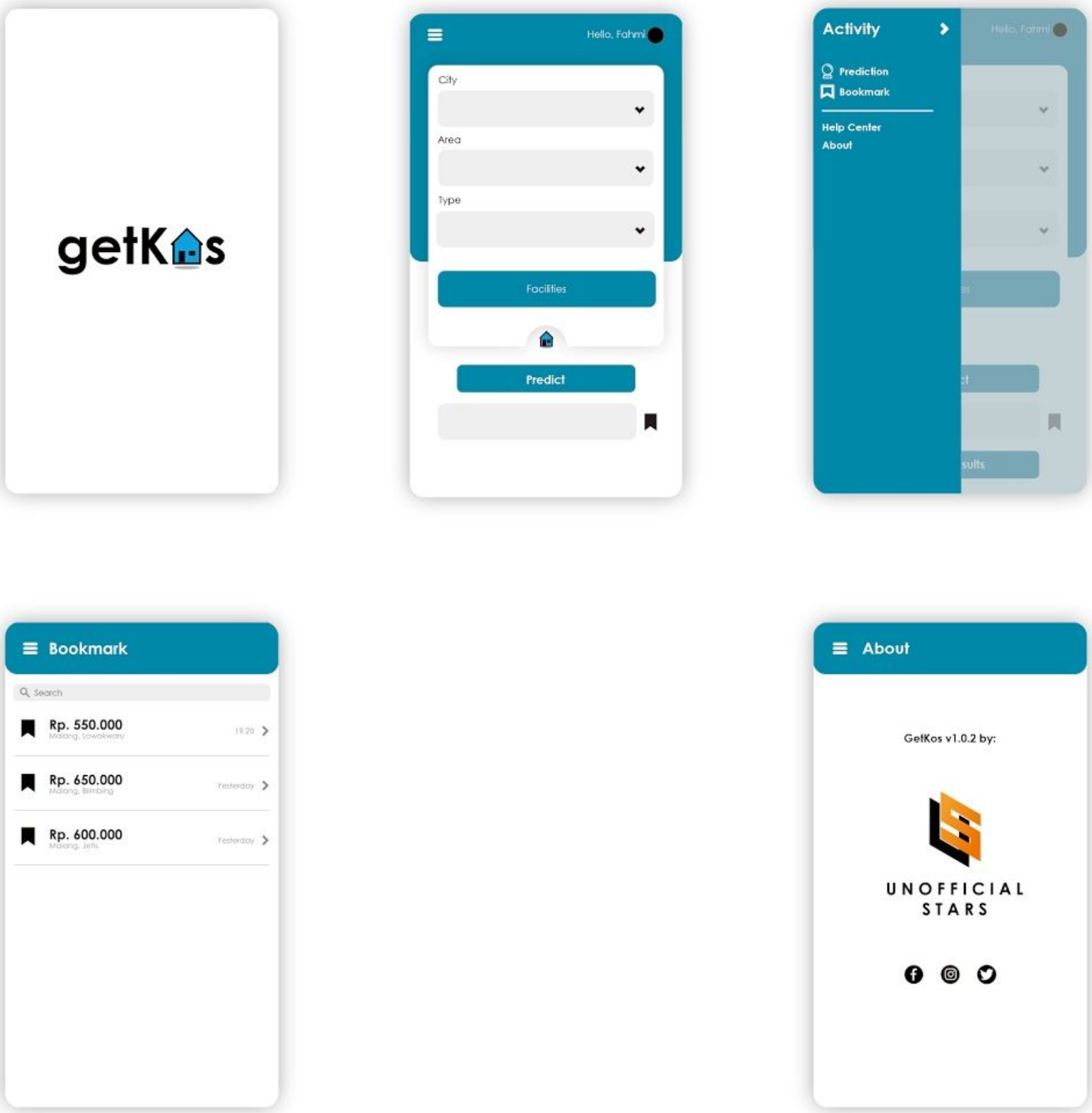


Figure 11. Display design of the GetKos application

In Figure 11 user interface of the application that we created, in general this application consists of 5 views, namely the splash screen, prediction screen, sidebar, bookmark screen, about screen and help center screen.

### 3.9 Android Apps Development

Implementation of the model for boarding house rent prediction price, we put it on the android platform. Which the development of this mobile app, we use several tools for development, testing and implementation. Includes:

a. IDE Android Studio

Android studio is an Integrated Development Environment that is used to develop applications on the android operating system. This IDE is based on Intellij IDEA.
````

Android studio offers many productivity-enhancing features for developing android based applications.

b. Genymotion

Genymotion is an android emulator that has quite complete features, from sensors and features to interact with the android virtual environment. In terms of testing, Genymotion is better than using a virtual machine in Android Studio. This is because the virtual machine is run separately from the Android Studio. So it's better to do debugging.

c. TensorFlow Lite

Tensorflow lite is a tool that helps developers to run tensorflow models on the mobile side, cropping systems and IoT devices. Through tensorflow lite, the device can perform inference or deduction with less latency and a smaller binary size, instead of using the usual tensorflow. In its implementation, TensorFlow Lite consists of an interpreter that acts as a model optimizer for certain hardware and a converter that converts the model into a form that is more efficient for use.

## 4. Result and Discussion

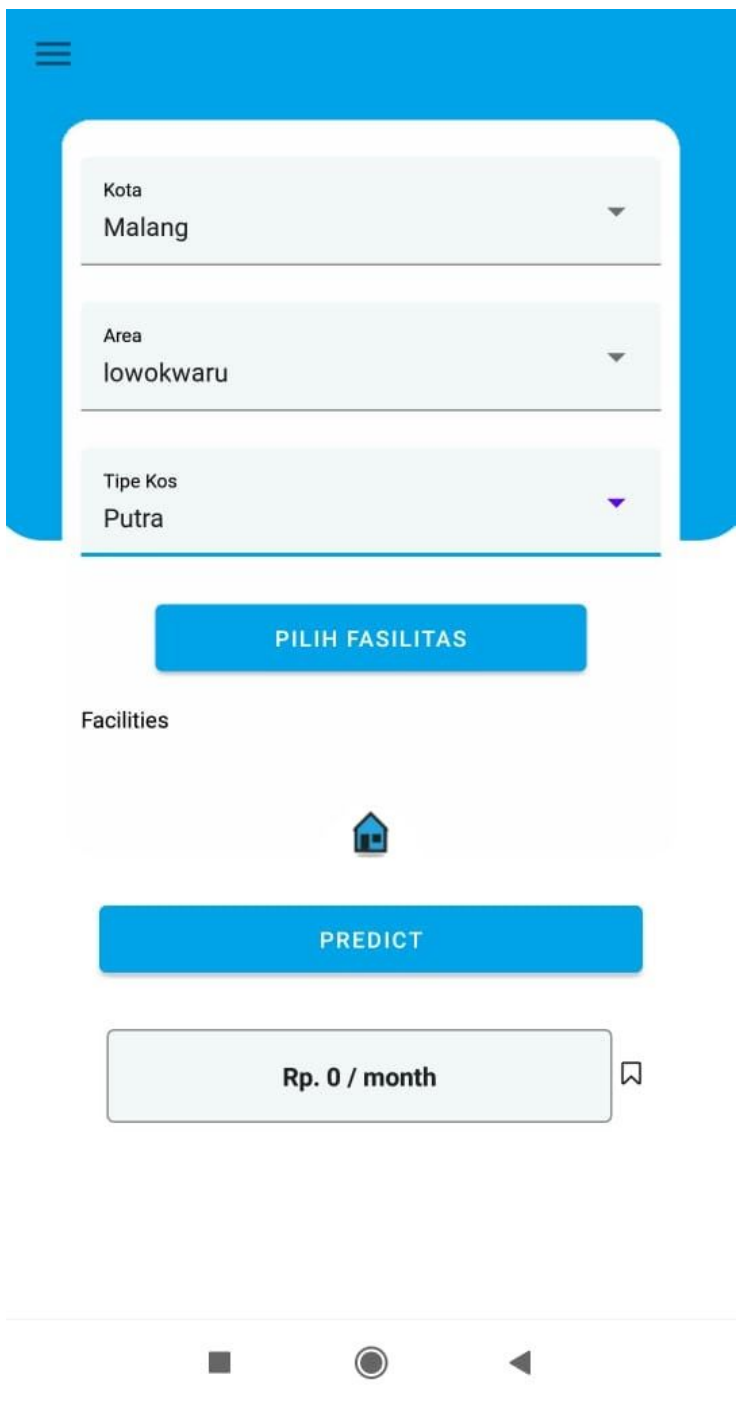

Figure 12. Step I

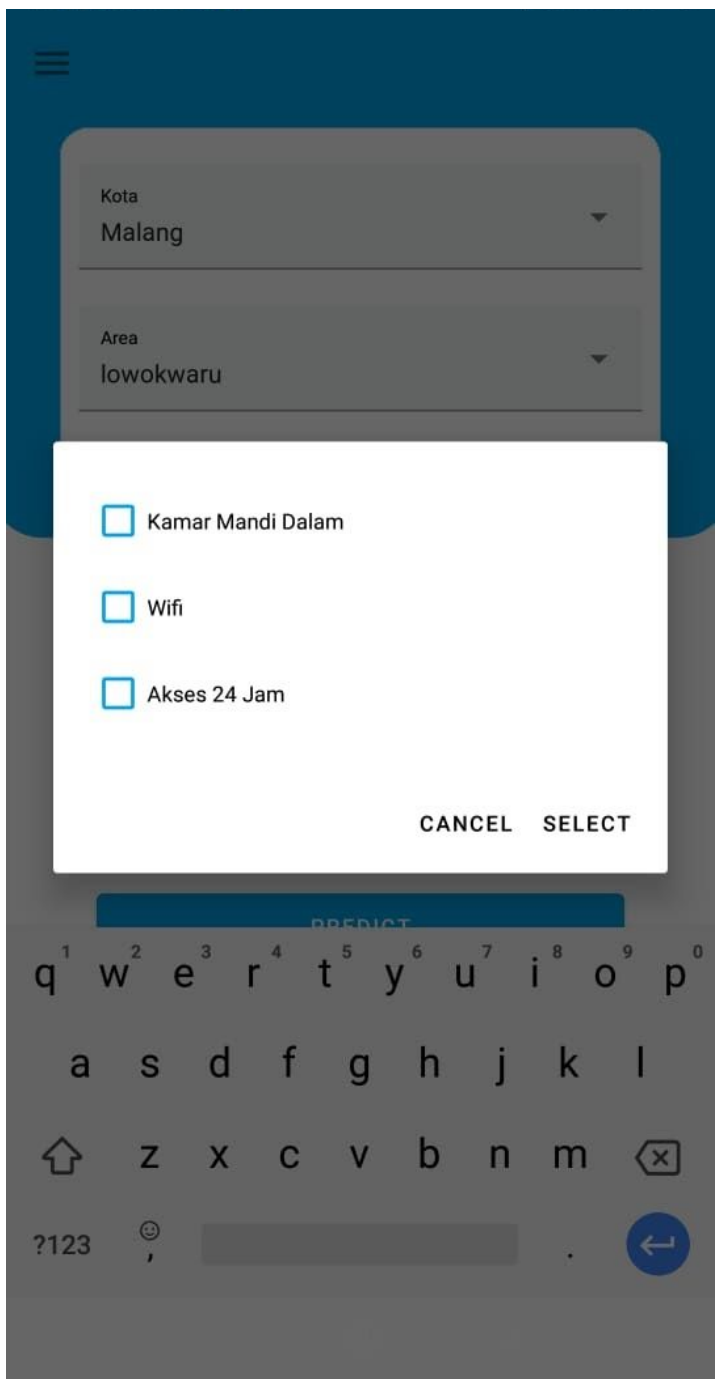

Figure 13. Step II

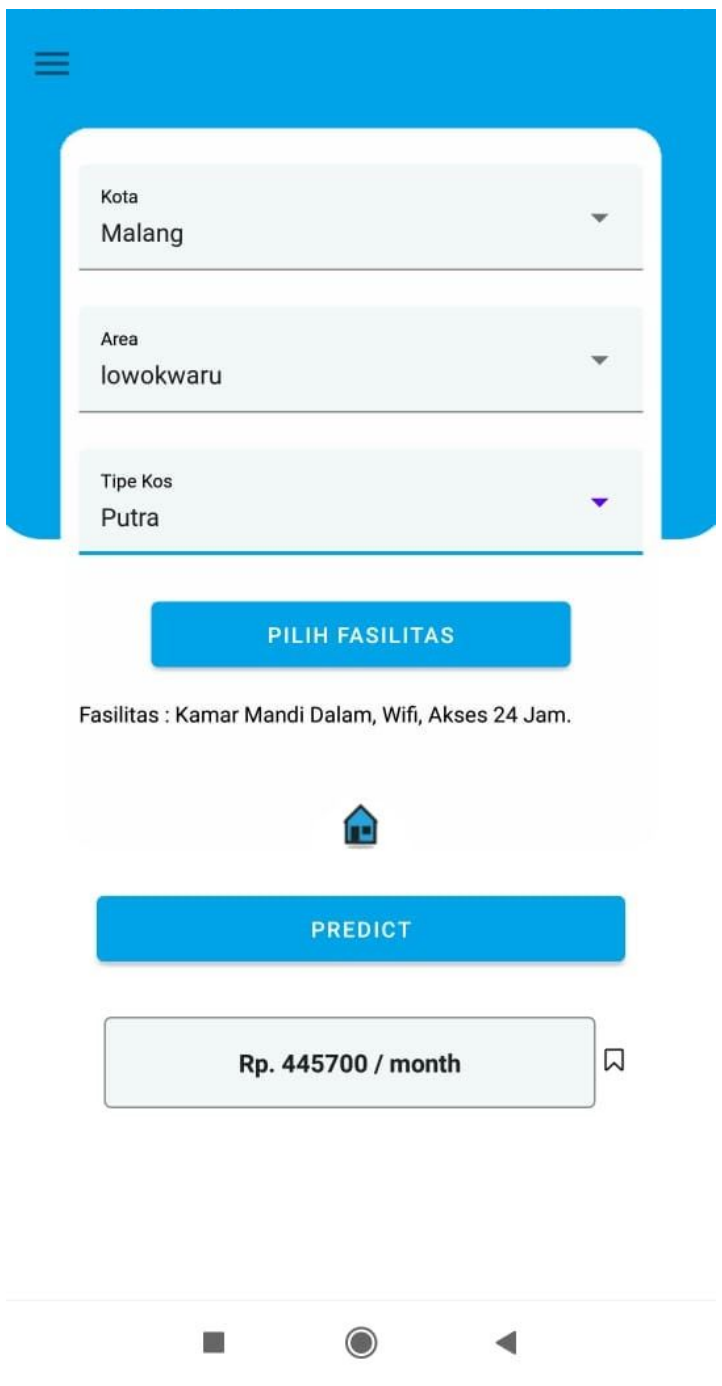

Figure 14. Step III

This is the result of the implementation on the mobile phone app from all the stages described in the previous stages. In Figure 12 is step I, which is to enter the input you want to know the price of the boarding house rent in a certain area by entering data in the form of city, area, boarding type, and in Figure 13 is step II entering the desired facilities. Then in Figure 14 is step III which can be seen the results of the prediction of the data obtained.

## 5. Conclusion and Future Work

We successfully implemented it into a mobile phone with the most ideal price using the neural network regression model and we succeeded in predicting boarding prices based on predetermined variables. For the next development, you can add the amount of data, both from the number of cities and the overall data. In addition, in the future people who have boarding houses can also enter data independently and the system can train their own data, so that the prediction results obtained can be updated automatically and have more consistent updates.